\documentstyle[12pt,scipp,psfig]{article}
\catcode`@=11
\def\section{\@startsection{section}{0}{\z@}{-3.5ex plus -1ex minus
 -.2ex}{2.3ex plus .2ex}{\large\bf}}
\@addtoreset{equation}{section}
\catcode`@=12
\renewcommand{\theequation}{{\rm \thesection.\arabic{equation}}}

\newcommand{\appendixA}{\setcounter{equation}{0}
\def\theequation{\rm{A}.\arabic{equation}}\section*}

\catcode`\@=11
\def\marginnote#1{}
\def\ifmath#1{\relax\ifmmode #1\else $#1$\fi}

\def\pri{^{\, \prime }}

\def\hl{h^0}
\def\mhl{m_{\hl}}

\def\rta{\rightarrow}
\def\bold#1{\setbox0=\hbox{$#1$}%
     \kern-.025em\copy0\kern-\wd0
     \kern.05em\copy0\kern-\wd0
     \kern-.025em\raise.0433em\box0 }

\def\GENITEM#1;#2{\par\vskip6pt \hangafter=0 \hangindent=#1
   \Textindent{$ #2$ }\ignorespaces}

\newcount\hour
\newcount\minute
\newtoks\amorpm
\hour=\time\divide\hour by60
\minute=\time{\multiply\hour by60 \global\advance\minute by-
\hour}
\edef\standardtime{{\ifnum\hour<12 \global\amorpm={am}%
    \else\global\amorpm={pm}\advance\hour by-12 \fi
    \ifnum\hour=0 \hour=12 \fi
    \number\hour:\ifnum\minute<100\fi\number\minute\the\amorpm}}
\edef\militarytime{\number\hour:\ifnum\minute<100\fi\number\minute}
\def\draftlabel#1{{\@bsphack\if@filesw {\let\thepage\relax
  \xdef\@gtempa{\write\@auxout{\string
    \newlabel{#1}{{\@currentlabel}{\thepage}}}}}\@gtempa
    \if@nobreak \ifvmode\nobreak\fi\fi\fi\@esphack}
     \gdef\@eqnlabel{#1}}
\def\@eqnlabel{}
\def\@vacuum{}
\def\draftmarginnote#1{\marginpar{\raggedright\scriptsize\tt#1}}
\def\draft{\oddsidemargin -.5truein
        \def\@oddfoot{\sl preliminary draft \hfil
        \rm\thepage\hfil\sl\today\quad\militarytime}
        \let\@evenfoot\@oddfoot \overfullrule 3pt
        \let\label=\draftlabel
        \let\marginnote=\draftmarginnote

\def\@eqnnum{(\theequation)\rlap{\kern\marginparsep\tt\@eqnlabel}%
\global\let\@eqnlabel\@vacuum}  }
\def\preprint{\twocolumn\sloppy\flushbottom\parindent 1em
        \leftmargini 2em\leftmarginv .5em\leftmarginvi .5em
        \oddsidemargin -.5in    \evensidemargin -.5in
        \columnsep 15mm \footheight 0pt
        \textwidth 250mm      \topmargin  -.4in
        \headheight 12pt \topskip .4in
        \textheight 175mm
        \footskip 0pt

\def\@oddhead{\thepage\hfil\addtocounter{page}{1}\thepage}
        \let\@evenhead\@oddhead \def\@oddfoot{} \def\@evenfoot{}
}
\def\titlepage{\@restonecolfalse\if@twocolumn\@restonecoltrue\onecolumn
     \else \newpage \fi \thispagestyle{empty}\c@page\z@
        \def\thefootnote{\fnsymbol{footnote}} }
\def\endtitlepage{\if@restonecol\twocolumn \else  \fi
        \def\thefootnote{\arabic{footnote}}
        \setcounter{footnote}{0}}  
\catcode`@=12
\relax
\def\ifmath#1{\relax\ifmmode #1\else $#1$\fi}
\def\half{\ifmath{{\textstyle{1 \over 2}}}}
\def\be{\begin{equation}}
\def\ee{\end{equation}}
\def\bea{\begin{eqnarray}}
\def\eea{\end{eqnarray}}
\def\simlt{\stackrel{<}{{}_\sim}}
\def\simgt{\stackrel{>}{{}_\sim}}

\def\NPB#1#2#3{{\sl Nucl.~Phys.} {\bf{B#1}} (19#2) #3}
\def\PLB#1#2#3{{\sl Phys.~Lett.} {\bf{B#1}} (19#2) #3}
\def\PRD#1#2#3{{\sl Phys.~Rev.} {\bf{D#1}} (19#2) #3}
\def\PRL#1#2#3{{\sl Phys.~Rev.~Lett.} {\bf{#1}} (19#2) #3}
\def\ZPC#1#2#3{{\sl Z.~Phys.} {\bf C#1} (19#2) #3}
\def\PTP#1#2#3{{\sl Prog.~Theor.~Phys.} {\bf#1}  (19#2) #3}

\def\mst11{m_{\;\widetilde{t}_{1}}}

\def\mst22{m_{\;\widetilde{t}_{2}}}
\def\mst12{m_{\;\widetilde{t}_{1,2}}}

\def\msb11{m_{\;\widetilde{b}_{1}}}
\def\msb22{m_{\;\widetilde{b}_{2}}}
\def\msb12{m_{\;\widetilde{b}_{1,2}}}

\def\mtilde2{\widetilde{m}^{2}}

\relax

\begin{document}
\topmargin-2.5cm
%
\begin{titlepage}
\begin{flushright}
CERN-TH/95-311\\
SCIPP 95/44 \\
hep--ph/9512446 \\
\end{flushright}
\vskip 0.3in
\begin{center}{\Large\bf Four-Generation Low Energy Supersymmetry \\[2pt]
with a Light Top Quark}
\vskip .5in
{\bf M. Carena},
{\bf H.E. Haber}%
\footnote{Permanent address:
Santa Cruz Institute for Particle Physics, University of California,
Santa Cruz, CA 94064  USA.} and {\bf C.E.M. Wagner}
\vskip.35in
CERN, TH Division, \\ 
CH--1211 Geneva 23, Switzerland\\
\end{center}
\vskip1.3cm
\begin{center}
{\bf Abstract}
\end{center}
\begin{quote}
A supersymmetric model with four generations is proposed,
in which the top quark is approximately degenerate
in mass with the $W^{\pm}$ gauge boson, $m_t\simeq m_W$,
leading to values of $R_b$ in better agreement with the
present experimental data than in the Standard Model.
The model shares many of the good features of the minimal
supersymmetric extension of the Standard Model (MSSM), such as
the unification of gauge and Yukawa couplings at a common 
high-energy scale.   The model differs from the MSSM by re-interpreting
the  Tevatron ``top-quark'' events as the
production of the fourth generation quark $t\pri$, 
which decays dominantly to $bW^+$.
The top quark decays primarily into supersymmetric particles,
$t\rta\widetilde t\widetilde\chi^0_1$, with $\widetilde t\rta
c\widetilde\chi^0_1$, thereby evading previous searches.
Light supersymmetric particles are predicted to lie
in the mass range between 25 and 70 GeV,
which together with the fourth generation leptons provide
a rich spectrum of new physics which can be probed at LEP-2 and the
Tevatron. 
\end{quote}
\vskip2.cm

\begin{flushleft}
CERN-TH/95-311\\
December 1995 \\
\end{flushleft}

\end{titlepage}
\setcounter{footnote}{0}
\setcounter{page}{0}
\newpage
%
\section{Introduction}

Recently, the CDF and D0 Collaborations have announced the discovery
of the top quark at the Tevatron \cite{topquark}, with a measured
mass of $m_t=176\pm 8\pm 10$~GeV and
$m_t=199^{+19}_{-21}\pm 22$~GeV, respectively.  Both measurements
are in excellent agreement with the top quark mass deduced by the LEP
global analysis of precision electroweak
measurements.  The most recent numbers quoted by the LEP Electroweak
Working Group from a Standard Model fit with a Higgs mass of
300~GeV are: $m_t=170\pm 10^{+17}_{-19}$~GeV
[$180^{+8}_{-9}\,^{+17}_{-20}$~GeV] based on the global average of the
data from the four LEP detectors and excluding [including] data from the
SLD Collaboration \cite{lepglobal}, where the last positive and negative
error quoted corresponds to alternative Higgs mass choices of
1~TeV and 60~GeV, respectively.  The LEP determination of $m_t$
derives from
the sensitivity of electroweak observables in $Z$ decays
to virtual top quark exchange, which
enters in two distinct ways.  First, top quark loops
in gauge boson self-energies (the so-called oblique corrections)
can directly affect the properties of the $Z$ and $W^\pm$.  The most
important oblique correction is the top-quark contribution to
the electroweak $\rho$ parameter \cite{rhoparm}, which is given by
$\rho=1+\Delta\rho$, where
$\Delta\rho\simeq 3G_F m_t^2/8\pi^2\sqrt{2}$.
Second, top-quark loops can contribute to three-point (and higher)
interactions (these are the non-oblique corrections).  The most
significant non-oblique correction to $Z$ decays arises in the
vertex correction to $Z\rta b\bar b$, which
is also quadratically sensitive to the top quark mass.
The size of the oblique
corrections, which can be inferred from the present precision data
analysis, is in very good agreement with the Standard Model
theoretical expectations.  In contrast, the measured value of the
rate for $R_b\equiv \Gamma(Z\rta b\bar b)/\Gamma(Z\rta{\rm hadrons})$
lies more than three standard deviations
above the Standard Model prediction.

If the deviation in the observed value of $R_b$ holds under further
experimental scrutiny, then this would be a signal of new physics
beyond the Standard Model, arising (presumably) from the virtual
exchange of new particles.  In the literature, models of
low-energy supersymmetry have been examined in which $R_b$ is
slightly enhanced above the Standard Model 
prediction \cite{BF,Renard,ABF,Gordy,Joan,Stefan,Hollik,CW,topdec,jellis}.
In such models,
the global fit to the electroweak data is improved.  Note that
in order to improve the Standard Model fit, one must approximately
maintain the size of the Standard Model oblique corrections
while modifying the $Zb\bar b$ interaction.  In the minimal
supersymmetric extension of the Standard Model (MSSM), this
is possible if one takes large [small] values of the
mass parameters of the scalar super-partners of the
left [right] handed top quark, and small values of the Higgs superfield
mass parameter $\mu$.

In this paper, we take an alternative approach.  If one considers
the $R_b$ measurement alone within the context of the Standard Model,
one would conclude that $m_t \simlt m_W$.  To avoid conflict with
the measurement of $m_t$ at the Tevatron (by direct means) and at
LEP (by indirect means) as described above, one must invoke new
physics to explain the CDF and D0 ``top-quark'' events and new
virtual effects to generate the oblique radiative corrections
observed in precision electroweak experiments.  In addition, one
must explain why such a light top-quark has escaped detection in
previous experiments at hadron colliders.  Surprisingly, a model
with such a light top-quark exists which is not in conflict with
present experimental observations.   The model consists of the
MSSM extended to four generations, with a light top-squark such
that $t\rta\widetilde t\widetilde\chi^0_1$ with nearly 100\%
branching ratio (where $\widetilde\chi^0_1$ is the lightest
neutralino, assumed to be the lightest supersymmetric particle).
The CDF and D0 ``top-quark'' events are identified as $t\pri\rta
bW^+$, and the oblique radiative corrections observed in
precision electroweak data are approximately saturated by loops
of third and fourth generation quarks and squarks.

In section 2 we review in more detail the top-quark mass dependence
of precision electroweak observables, and discuss the implications
of the recent measurement of $R_b$ (and the corresponding measurement
of $R_c$).  We discuss some of the requirements for the
viability of the light top quark explanation of the observed
discrepancy in $R_b$.
In section 3 we analyze the restrictions on the fourth
generation masses due to a variety of phenomenological
and theoretical constraints.   By imposing the requirement
that none of the couplings of the model become nonperturbative below
the grand unification (GUT) scale, we find that all fourth
generation fermion--Higgs Yukawa couplings approximately
unify at the GUT scale while approaching quasi-infrared fixed
point values at the electroweak scale.
In sections 4--6 we provide a detailed phenomenological profile of
our model, and discuss the constraints on its parameters,
with particular attention given to the implications of the
precision electroweak data and flavor changing processes.   
Section 7 summarizes the successes
and the shortcomings of our model.  Some technical details are
included in the Appendix.  A preliminary version of this work was
presented in Ref.~\cite{Moriond}.

\section{The top quark mass from precision electroweak data}
\medskip

Experiments at LEP and SLC measure more than fifteen separate
electroweak observables in $Z$ decay events.  Global fits of
these observables exhibit a remarkable consistency with Standard
Model expectations, with two notable exceptions.
Defining $R_q\equiv\Gamma(Z\rta
q\bar q)/\Gamma(Z\rta{\rm hadrons})$, with
$q = b,c$, the LEP Electroweak Working Group global fit 
yields \cite{lepglobal}
\begin{eqnarray}
R_b=\cases{0.2219\pm 0.0017,&LEP/SLC global fit ;\cr
0.2156,&Standard Model prediction,\cr}
\label{zbbnumber1}
\end{eqnarray}
which is a $3.7\sigma$ discrepancy, and
\begin{eqnarray}
R_c=\cases{0.1543\pm 0.0074,&LEP/SLC global fit ;\cr
0.1724,&Standard Model prediction,\cr}
\label{zccnumber1}
\end{eqnarray}
which is a $2.5\sigma$ discrepancy.  
One other LEP measurement relevant to this discussion is the
$\alpha_s(m_Z)$ determination \cite{Stefan,als,gordy2,Joan2}
from the total hadronic width of the 
$Z$.  Based on the measurement of 
$R_\ell\equiv\Gamma_{\rm had}/\Gamma_{\ell\ell}$, the Standard Model fit
yields \cite{lepglobal}
$\alpha_s(m_Z)=0.126\pm 0.005\pm 0.002$ (where the last error
quoted corresponds to varying the Higgs mass from 60~GeV to 1~TeV).
It is interesting to 
note that the LEP determination of $\alpha_s(m_Z)$ tends to be 
somewhat higher than the extrapolated value of $\alpha_s(m_Z)$
obtained from lower energy measurements.

One must be very careful in interpreting the $R_b$ and $R_c$
discrepancies from Standard Model expectations.
The experimental procedures that identify $b$ and $c$ quarks in
$Z$ decays are difficult and prone to large
systematic errors.  Regarding the $R_c$ measurement, we note first
that the quoted error is larger, and the
statistical significance of the deviation from the Standard Model
prediction is smaller, than those of $R_b$.  Second, 
it is very difficult to imagine new physics processes
that could produce such a large deviation in $R_c$.
To understand the latter point, suppose new
physics were responsible for both the $R_b$ and $R_c$ discrepancies.
Because $R_b+R_c$ is less than the Standard Model prediction,
one is led to a distressing conclusion.
Since $\alpha_s(m_Z)$ is determined from the experimentally
measured $\Gamma_{\rm had}$, a negative 
contribution of new physics to this quantity implies that the 
perturbative QCD corrections to $\Gamma_{\rm had}$ should be larger.
That is, $\alpha_s(m_Z)$ determined from $\Gamma_{\rm had}$ would
be larger than the number quoted above, in significant conflict
with the low-energy determinations.  This argument implicitly
assumes that there are
no new physics contributions to $R_q$ where $q$
is a light quark.   On the other hand, there is no known source
of new physics that can modify $R_q$ sufficiently to compensate
the deficit in $R_b+R_c$ to avoid the above conclusion.
Thus, we are inclined to discount the measured value of $R_c$ above,
and assume that its true value is close to the Standard Model
expectation.  Should we discount the measured value of $R_b$ as
well?  Further experimental analysis is required to
clarify the situation.  Here, we simply note that if new physics
contributes to $R_b$ but not to $R_c$ and $R_q$, then because
$R_b$ is above the Standard Model prediction, $\alpha_s(m_Z)$
determined from $\Gamma_{\rm had}$ will be {\it lower}, 
potentially closer to the low energy 
$\alpha_s$ determinations \cite{Stefan,als,gordy2,Joan2}.

Henceforth, we assume that $R_c$ is given by its Standard Model
prediction.  In the experimental determination of $R_b$, there
is some contamination of $c\bar c$ events in the $b\bar b$
sample that must be subtracted.  This subtraction depends on
the value of $R_c$ assumed.  Fixing $R_c$ to its Standard Model
value, a slightly smaller value of $R_b$ is found 
by the Electroweak working group compared to
the value quoted above:
\begin{eqnarray}
R_b=0.2205 \pm 0.0016,\qquad\qquad {\rm LEP/SLC~global~fit,}
\label{zbbnumber2}
\end{eqnarray}
roughly a three standard deviation discrepancy
from the Standard Model prediction.

The $Z b \bar{b}$ vertex corrections can also significantly affect the 
left-right $b \bar{b}$ asymmetry ${\cal{A}}_b$, 
\begin{equation}
{\cal{A}}_b = \frac{ g_L^2 - g_R^2}{g_L^2 +g_R^2},
\end{equation} 
where $g_L$ ($g_R$) are  the couplings of the left (right)
handed bottom quarks to the $Z$. The corrections to $R_b$ and 
${\cal{A}}_b$ can be parameterized as a function of the corrections
to the left-- and right--handed bottom quark vertices,
\begin{equation}
\frac{\delta {\cal{A}}_b}{{\cal{A}}_b} = \frac{4 f_R f_L}
{f_L^4 - f_R^4} \left[f_R \delta g_L - f_L
\delta g_R \right],
\end{equation}
\begin{equation}
\frac{\delta R_b}{R_b} = \frac{2 \; (1-R_b)}{f_L^2 + f_R^2}
\left[f_R \delta g_R  + f_L \delta g_L \right],
\end{equation}
where $f_L = 1/2 -  \sin^2\theta_W(m_Z)/3$
and $f_R = - \sin^2\theta_W(m_Z)/3$ are the tree level
couplings of the left and right handed bottom quarks to
the $Z$. The dominant,
top quark mass dependent, one loop $Zb\bar{b}$ vertex corrections
affect only the left--handed bottom quark coupling to the $Z$, 
\begin{equation}
\delta g_L = - \frac{\alpha}{16 \pi \sin^2\theta_W} m_t^2.
\label{dglst}
\end{equation}
The large difference between the values of $f_L$ and $f_R$
implies that for $\delta g_R = 0$,
$\delta R_b/R_b \simeq 11.5 \; \delta {\cal{A}}_b/{\cal{A}}_b$.
Moreover, the current  determination of ${\cal{A}}_b$ at SLC 
is still subject to large experimental errors \cite{lepglobal}
\begin{eqnarray}
{\cal{A}}_b = \cases{0.841 \pm 0.053 &LEP/SLC global fit; \cr
0.935   ,&Standard Model prediction.\cr}.
\label{Abnumber}
\end{eqnarray}
Therefore,  ${\cal{A}}_b$ does not provide at present any significant
constraint on the value of the top quark mass.

Barring additional experimental problems, and even after taking
into account the possible systematic errors due to an incorrect
charm background subtraction, 
the discrepancy in $R_b$ quoted in Eq.~(\ref{zbbnumber2}) is
large enough as to raise concerns about
the Standard Model interpretation of the data. As noticed above,
if one extracts the prediction of the top quark mass
from the $R_b$ measurement alone, one would
conclude that the top quark is
much lighter than the value inferred from the global LEP
fit to the data or from the
direct Tevatron measurements. Of course,
with such a light top quark, one must address the obvious questions
of why such a top quark has not yet been discovered, and what is the
particle recently announced by the
CDF and D0 Collaborations that is observed to
decay into $bW^+$.

If the top quark were
sufficiently light, then $W^+\rta t\bar b$ would be kinematically
allowed; this would modify the total width of the $W^\pm$.  But
$\Gamma_W$ can be measured at hadron colliders indirectly by
studying the ratio of production cross section times
leptonic branching ratio of the
$W^{\pm}$ and $Z$.  The most recent analysis of this kind, reported
by the CDF collaboration \cite{wwidthcdf}, 
finds $m_t>62$~GeV at 95\% CL.
(The most recent D0 data \cite{wwidthdzero}
implies a similar bound.  A global average of all hadron collider data
would raise this limit by a few GeV.) 
Direct searches for the top quark at hadron colliders assume that an
observable fraction of top quark decays results in a lepton
in the final state.  For example,
the D0 collaboration ruled out the mass range
$m_W+m_b\simlt m_t<131$~GeV, assuming that the decay $t\rta bW^+$ is
not unexpectedly suppressed \cite{dzerotop}.
Previous top quark searches at hadron
colliders were able to close the window between 62 GeV and
$m_W + m_b \simeq 85$ GeV,
assuming that $t\rta bW^\star$ is the dominant top-quark decay mode.
However, in the latter case, the final state is three-body since $W^\star$
is virtual.  If the top quark possessed any two-body decay modes (due
to new physics processes), and if these modes rarely produced
leptons, then a top quark in this mass region could have escaped detection
in all previous top quark searches.

One possible scenario leading to two-body decay modes of the top quark
arises if one assumes the existence of a light charged
Higgs particle in the spectrum, $m_{H^\pm} < m_W$.
Recent results from the CDF Collaboration rule out a substantial
portion of the allowed parameter space \cite{toptoHiggs} based on a
direct search for $t\to bH^+$.  Additional constraints arise from
considering the contribution of virtual charged Higgs effects to heavy
quark decays.  In particular, one-loop charged Higgs exchange can 
significantly modify the predictions for rare bottom quark decay
rates.  Indeed, from the requirement of obtaining an acceptable
rate for the branching ratio ${\rm BR}(b \rightarrow s \gamma)$,
one can show that this scenario is ruled out \cite{Hou}, 
unless the enhancement of this branching ratio due to the presence of
a light charged Higgs boson is compensated by the contribution of
new particles appearing at the electroweak scale.

Thus, we turn to an alternative scenario, based on low-energy 
supersymmetry with a light top-squark,
in which the decay
$t\rta\widetilde t\widetilde\chi^0_1$ is kinematically allowed
(where $\widetilde t$ is the lightest top squark and
$\widetilde\chi^0_1$ is the lightest neutralino) \cite{rudaz}.  
Experimental searches for
both $\widetilde t$ and $\widetilde\chi^0_1$ place constraints on their
masses, but do not rule out the possibility of $M_{\tilde t}+
M_{\widetilde\chi_1^0}<m_W$.
In particular, the LEP neutralino and chargino searches \cite{pdg}
imply a limit on the lightest neutralino mass which typically lies
between 20 and 25 GeV (although lighter neutralino masses cannot be
ruled out if $\tan\beta$ is very
close to one).  Using this result and the limits on the top squark
mass from searches at LEP and at the Tevatron \cite{stopsearch},
one finds that the mass region
50 GeV $\simlt M_{\tilde t}\simlt 60$~GeV (with
$M_{\widetilde\chi_1^0}\gsim 20$~GeV) cannot be excluded.%
\footnote{The most recent LEP run, at $\sqrt{s} =
136$ GeV, is not sensitive to this range of masses in the case of maximal
top squark mixing (the latter is a necessary feature of our model).}

To be definite, we choose $m_t\simeq m_W$, $M_{\tilde t}\simeq 50$~GeV and
$M_{\widetilde\chi^0_1}\simeq 25$~GeV.  Then, the dominant decay chain is
$t\rta\widetilde t\widetilde\chi^0_1$ followed by
$\widetilde t\rta c\widetilde\chi^0_1$ through a one-loop
process \cite{rudaz},  which rarely produces a hard isolated lepton.
Hence, these events would not have been detected at hadron colliders.
But, now we must
reconsider  the recent CDF and D0 discoveries and the LEP
``measurement'' of $m_t$.  As anticipated in section 1,
we propose to account for these results
by introducing a fourth generation of quarks (and leptons) plus
their supersymmetric partners.   Then, $t\pri\rta bW^+$ can be the
source of the CDF and D0 events, while the effects of
the third and fourth generation quarks and squarks contributing to
the oblique corrections are large enough
to be consistent with LEP precision electroweak
data.

\section{Fourth generation fermion masses and unification}
\medskip

For the $t\pri$ to be consistent with the CDF and
D0 ``top-quark'' events, its mass should be close to the
world average, $m_{t\pri} = 180 \pm 12$ GeV.
We assume that the $t\pri$ mass is
given by $m_{t\pri} \simeq 175$ GeV. Moreover, in order to
produce the observed Tevatron signal, the dominant $t\pri$ decay
must be $t\pri\rta bW^+$, in which the generation number changes
by one unit.  This means that two body generation number conserving
decays, such as $t\pri\rta b\pri W^+$ or the supersymmetric
decay modes $t\pri\rta\widetilde{b\pri}\widetilde\chi^\pm$ and
$t\pri\rta\widetilde{t\pri}\widetilde\chi^0$, must be kinematically
forbidden.  Of course, three-body generation number conserving decays
will be possible ({\it e.g.}, $t\pri\rta b\pri W^*$, where the 
virtual $W^\star$ decays in the usual way to a fermion antifermion pair).
It follows that the $t\pri$--$b$ mixing angle
($V_{t\pri b}$) must not be too small; otherwise the three-body
generation conserving decays will dominate.  For example, we find:
\begin{equation}
{\Gamma(t\pri\rta b\pri W^\star)\over\Gamma(t\pri\rta bW)}
={9G_Fm_{t\pri}^2|V_{t\pri b\pri}|^2
\over \pi^2\sqrt{2}|V_{t\pri b}|^2}
\int_0^{1-2\sqrt{x}+x}\,{z(1-z+x)\sqrt{(1-z+x)^2-4x}\over
\left(1-{zm_{t\pri}^2/m_W^2}\right)^2}\,dz\,,
\label{gammarats}
\end{equation}
where $x\equiv m_{b\pri}^2/m_{t\pri}^2$.
Since the rate of the CDF and D0 ``top-quark'' events is consistent
with the QCD prediction for $t\bar t$ production under the assumption
that ${\rm BR}(t\rta bW^+)\simeq$ 1, a reinterpretation of these events as
$t\pri \bar {t\pri}$ production (followed by $t\pri\rta bW^+$)
requires ${\rm BR}(t\pri\rta bW^+)$ to be near 1.  At present, there
are no relevant constraints on the mixing parameter $V_{t\pri b}$.
We assume that $V_{t\pri b}$
lies between $V_{cb}=0.04$ and $V_{us}=0.2$,
although even larger values may be 
allowed in some circumstances \cite{Pakvasa}.
For definiteness, we
choose $|V_{t\pri b}/V_{t\pri b\pri}|=0.1$.

In Figure 1, we show the dependence of the branching ratio
 ${\rm BR}(t\pri\rta bW^+)$ on the $b\pri$ mass. We see that
there is a sharp increase of this branching ratio as soon as
$m_{b\pri} + m_W \simgt m_{t\pri}$.
Then, if we require ${\rm BR}(t\pri\rta bW^+)
\simgt 0.75$ for the chosen value of
$V_{t\pri b}$, it follows that 
$m_{b\pri}\simgt 105$~GeV.  Such a large value
of the fourth generation down-type quark mass implies that the custodial
SU(2) symmetry breaking is softer than in the three generation
Standard Model case. As we shall discuss in section 5,
this has important consequences for
the value of the $\rho$-parameter. In particular,
values of the $b\pri$ mass too close to 175 GeV would immediately
imply a suppression of the one-loop contribution to the
$\rho$-parameter, which is
in conflict with the present precision electroweak
data. Hence, the $b\pri$ mass
should either be larger than the $t\pri$ mass or
close to its experimental lower bound,
$m_{b\pri} \simeq 105$ GeV. Having introduced a fourth generation
of quarks, the simplest way to avoid electroweak gauge anomalies is
to introduce a fourth generation of sequential leptons.  In order to
be consistent with present LEP-1 bounds, their masses
must be larger than 45 GeV (see discussion below).  

\begin{figure}[htb]
\centering
\centerline{
\psfig{file=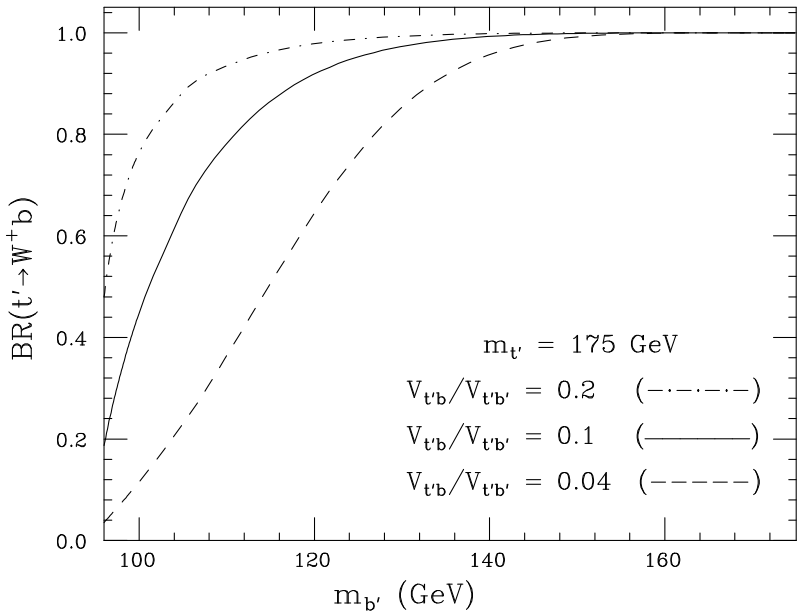,height=8cm,angle=90}}
\caption[0]{
Branching fraction ${\rm BR}(t\pri \rightarrow b W^+)/BR(t\pri
\rta b\pri W^+)$
as a function of
the fourth generation bottom mass $m_{b\pri}$.}
\end{figure}

In the absence of
any assumption about the nature of the physics at high energy scales, it
is simple to construct a four-generation model, which fulfills all
the relevant experimental bounds, by simply pushing up the
$b\pri$ quark mass above the $t\pri$ quark mass.
In contrast, we find that
our four-generation model is remarkably well-constrained
if we demand that none of the gauge or Yukawa couplings
blow up below the GUT scale (or, in the absence of grand unification,
the Planck scale).  In the latter case, the parameters
of the model are almost uniquely determined, and the model is on the
verge of being ruled out (or poised for experimental discovery).

One of the strongest motivations for low-energy
supersymmetric extensions of the Standard Model comes
from the fact that such models can be realized as effective low energy
descriptions of theories in which all known low-energy
gauge interactions unify at a high energy scale. In the MSSM,
the values of the low energy gauge
couplings and third generation fermion masses are consistent with
the unification of gauge and Yukawa couplings at a single high
energy scale $M_{\rm GUT} \simeq 2 \times 10^{16}$~GeV \cite{SUSYG}. 
A fourth generation of fermions and their supersymmetric partners
can be accommodated in complete SU(5) superfield multiplets,
{\bf 10} + ${\bf{\overline{5}}}$, and does not affect the successful
$\sin^2\theta_W(m_Z)$ prediction coming from the gauge coupling
unification.  The two-loop running of the gauge couplings is
only slightly
modified by the presence of the fourth generation, leading to a
value of $\alpha(M_{\rm GUT})\simeq 0.08$ for the unified gauge
coupling and a slightly larger value of the unification scale,
$M_{\rm GUT} \simeq 5 \times 10^{16}$  GeV.  Moreover, taking
$\alpha_{em} (m_Z)$ and $\sin^2\theta_W(m_Z)$ as input
parameters, the two loop corrections enhance slightly the
predicted values of $\alpha_s(m_Z)$ with respect to the 
value obtained in the MSSM \cite{CPW}. 
This  enhancement is partially suppressed  by the effect of  
the large fourth generation Yukawa couplings, which happen to be
close to their fixed point values. In any case, as in the
MSSM, the predicted value of the strong gauge coupling,
$\alpha_s(M_Z) \simeq 0.13$, is somewhat high to
be consistent with the values of the strong gauge coupling
extracted from LEP within the assumptions of this model. 
However, it is enough to assume the presence of small
threshold corrections at the grand unification
scale to produce a predicted value of $\alpha_s(M_Z)$ that is
consistent with the present experimental data.

Consider next
the requirement of perturbative consistency of the Yukawa sector
of the theory up to scales of order $M_{\rm GUT}$.
The low energy Higgs-$b\pri$ quark
Yukawa coupling is related to the running mass $m_{b\pri}$ by
\begin{equation}
h_{b\pri} = \frac{\sqrt{2} m_{b\pri}}{v \cos\beta} \; ,
\end{equation}
where $v=246$~GeV and $\tan\beta$ is the ratio of the two Higgs
vacuum expectation values. For a fixed value of $m_{b\pri}$
an upper bound on $\tan\beta$ may be hence obtained. 
For a physical mass  $m_{b\pri} \simgt 105$~GeV,
only small values of
$\tan\beta < 2$ are allowed. On the other hand, the Higgs-$t\pri$ quark
Yukawa coupling is related to its running mass by
\begin{equation}
h_{t\pri} =  \frac{\sqrt{2} m_{t\pri}}{v \sin\beta}.
\end{equation}
Consequently, low values of
$\tan\beta$ tend to push the $t\pri$ mass to
relatively low values, unless its Yukawa coupling is itself
close to its perturbativity bound, corresponding to the quasi-infrared
fixed point value of this quantity. Large values of $h_{t\pri}$
have the additional effect of
changing the renormalization group evolution of $h_{b\pri}$,
leading to lower values of $m_{b\pri}$ (or equivalently, lower
values of $\tan\beta$ at fixed $m_{b\pri}$). 
As a result, in
order to obtain values of the fourth generation $b\pri$
and $t\pri$ masses consistent with the limits quoted earlier,
$m_{t\pri} \simeq 175$ GeV, $m_{b\pri}
\simgt 105$ GeV, two properties must be fulfilled:\\
~\\
1. The $b\pri$ mass must be close to its lower bound.  
Remarkably, this
conclusion coincides with the one obtained from the considerations of
the fourth generation contributions to the $\rho$-parameter,
discussed in more detail in section 5. \\
~\\
2. The $b\pri$ and $t\pri$ Yukawa couplings must lie close to
their quasi infrared fixed point values \cite{irfps}.
~\\

Therefore, the $b\pri$ mass is essentially fixed by perturbativity 
arguments.  Moreover, since we are close to the infrared fixed point
values of the Yukawa couplings,
$\tan\beta$ is also fixed.  
Figure 2 shows the resulting fourth generation quark masses
as a function of $\tan\beta$ assuming that both Yukawa couplings
lie close to their fixed point values.
For $m_{t\pri} \simeq 175$~GeV and $m_{b\pri}=105$~GeV;
we obtain $\tan\beta\simeq 1.6$.
Our picture is consistent
with the unification of the $t\pri$ and $b\pri$ Yukawa
couplings at $M_{\rm GUT}$ and, unlike
the case of gauge coupling unification, the predicted low
energy values are  highly insensitive to possible moderate
threshold corrections at $M_{\rm GUT}$. 

\begin{figure}[htbp]
\centering
\centerline{
\psfig{file=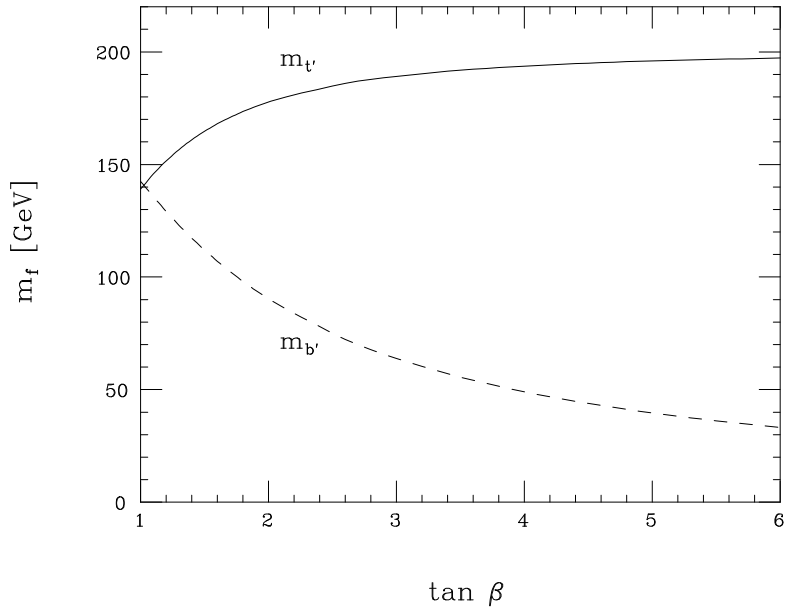,height=8cm,angle=90}}
\caption[0]{ Plot of the $b\pri$ and $t\pri$ quark
masses as a function of $\tan\beta$ assuming that both
lie close to their fixed point values.}
\end{figure}

The Tevatron may be able to rule out the existence of the $b\pri$
with mass $m_{b\pri}\simeq 105$~GeV (although the existing 
experimental limits
only rule out $b\pri$ quarks of mass up to 85~GeV \cite{pdg}).
If kinematically allowed, the decay $b\pri\rta\widetilde
t \widetilde\chi_1^-$ would be the dominant decay mode.  If disallowed,
there would be a competition between $b\pri\rta Wc$ (a change of two
generations) and $b\pri\rta W^\star t$ (a change of one generation, but
suppressed by three-body phase space).  One option is to choose
$|V_{b\pri c}|\ll |V_{t\pri b}|$ to remove the possibility of
$b\pri\rta Wc$.  Then, all $b\pri$ decays would result in
the final state $W^\star c\widetilde\chi_1^0\widetilde\chi_1^0$.
There are no published limits that exclude such a
$b\pri$.  However, a dedicated search at the Tevatron should be
able to discover or exclude such events.

\begin{figure}[htbp]
\centering
\centerline{
\psfig{file=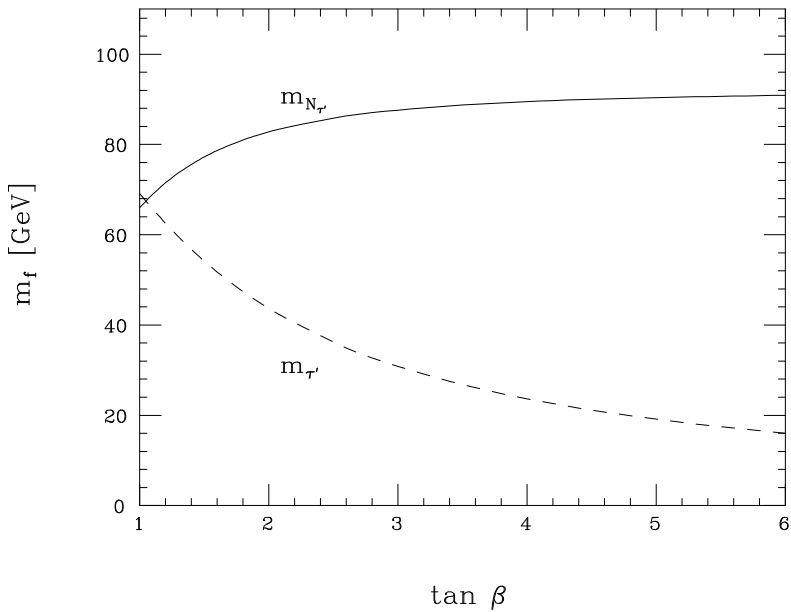,height=8cm,angle=90}}
\caption[0]{The same as Fig. 2 but for the $\tau$ and
the fourth generation neutrino masses.}
\label{tauneupfix}
\end{figure}

The fourth generation lepton Yukawa couplings are also 
strongly constrained by the experimental bound on the
lepton masses $m_{\ell^\prime} > 45$~GeV. When applied to
the neutrino mass, this bound implies that the right-handed
Majorana mass, if present, must be of order of the
electroweak scale.   In addition, the
bounds on the $\tau\pri$ lepton mass imply that its
Yukawa coupling must be larger than ${\cal O}(1)$ at the grand unification
scale. We find that the infrared fixed point value for
$m_{\tau\pri}$ is very close to the current experimental
bound on this quantity. Although there is more freedom in 
the $\tau\pri$--$N_{\tau^\prime}$ sector,
one can also add the requirement that both 
fourth generation Yukawa couplings take large
values at $M_{\rm GUT}$ (in order to maximize their masses). 
Figure~\ref{tauneupfix} shows the predicted values for the fourth
generation lepton masses assuming that the neutrino
$N_{\tau\pri}$ is a Dirac neutrino, and that the four
fourth-generation
Yukawa couplings are close to their fixed point values.
Then, for $\tan\beta \simeq 1.6$, the resulting lepton
masses are: $m_{\tau\pri}\simeq 50$~GeV and $m_{N_{\tau^\prime}}
\simeq 80$~GeV.  It is interesting to note that these lepton
masses are consistent with the unification of
all {\it four} fermion-Higgs Yukawa couplings at the GUT scale.
Moreover, these masses lie above the corresponding
bounds from LEP-1, but both are in the kinematic reach of LEP-2.

In general, the fourth generation neutrino mass matrix can contain
both an off-diagonal Dirac mass, $m_D$ and a right-handed diagonal
Majorana mass, $m_M$.  Thus, in the more general case,
the two fourth generation neutrino eigenstates are split by the
see-saw mechanism.  The lighter state has mass given by
\begin{equation}
m_{N_{\tau^\prime}} = \half\left[\sqrt{m_M^2+4m_D^2}-m_M\right]\,.
\end{equation}
The infrared fixed point limit of $m_D$ coincides
with that of $m_{N_{\tau\pri}}$ shown in Figure 3.  There is no
corresponding limit on the Majorana mass $m_M$.  For $m_D \simeq 80$ GeV, it is
easy to find the value of $m_M$ (roughly equal to
$m_D$) such that the light neutrino state becomes approximately
degenerate with the $\tau\pri$.  This observation implies that a
range of parameters exists where $N_{\tau\pri}$ is a few GeV
lighter than the $\tau\pri$.  Due to the near degeneracy of the two
states, it would be extremely difficult in this case
to detect the $\tau\pri$ at LEP-2
through its dominant decay $\tau\pri\to N_{\tau\pri}W^*$,
unless the $N_{\tau\pri}$ decayed inside the collider detector.   

\section{Phenomenological Restrictions and Implications}

In section 2, we noted that our model must possess a rather light
top-squark (stop) whose mass is about 50 GeV.  In order to achieve such a
small $M_{\widetilde t}$, there must be substantial
$\widetilde t_L$--$\widetilde t_R$ mixing.  $M_{\widetilde t}^2$
is given by the smallest eigenvalue of the stop squared-mass
matrix
\begin{eqnarray}
\pmatrix{M_Q^2+m_t^2+c_L m_Z^2 & m_t(A_t-\mu\cot\beta)\cr
 m_t(A_t-\mu\cot\beta)& M_U^2+m_t^2+c_R m_Z^2\cr}\,,
\label{stopmatrix}
\end{eqnarray}
where 
$M_Q$, $M_U$, and $A_t$ are soft-supersymmetry-breaking parameters,
$c_L\equiv ({1\over 2}-{2\over 3}\sin^2\theta_W)\cos2\beta$,
$c_R\equiv {2\over 3}\sin^2\theta_W\cos2\beta$, 
and $\mu$ is the supersymmetric Higgs mass parameter.  A light stop
implies that the off-diagonal terms in Eq.~(\ref{stopmatrix})
are of the same order as the
geometric average of the diagonal terms,
\begin{equation}
\left(m_{Q}^2 + m_t^2 \right)
\left(m_{U}^2 + m_t^2 \right) \simeq m_t^2 X_t^2
\label{tildeat}
\end{equation}
where $X_t\equiv A_t - \mu \cot\beta$.  This can be accomplished
either by taking $X_t$ rather large or by pushing
one of the soft supersymmetry breaking mass parameters
to low values. However, the latter possibility is severely constrained
by the constraints on the $\rho$-parameter and the bottom squark mass,
as discussed in section 5.

If there is large mixing in the third generation squark sector, one
would naively expect that the fourth generation squark mixing would be
large as well. If $X_{t\pri}$ were of order $X_t$,
the mixing in the fourth
generation squark sector would be too large, driving the smallest
eigenvalue of the ${\widetilde t}\pri_L$--${\widetilde t}\pri_R$
squared-mass matrix negative.  Remarkably,
due to the infrared fixed-point behavior of the fourth generation,
$A_{t\pri}$ is driven to a value that is independent of
its high energy value \cite{poketal}.
Roughly, $A_{t\pri}\simeq - 1.5 m_{1/2}$
where $m_{1/2}$ is the high-energy (GUT-scale) value of the gaugino Majorana
mass.  In contrast, due to the smaller values of the top quark Yukawa
coupling, the top squark mixing parameter $A_t$
is strongly dependent on its high
energy value and hence can be taken as an independent parameter.
Moreover, choosing $\mu$ negative and $A_t$ positive
enhances the third generation
squark mixing while somewhat suppressing the fourth generation
squark mixing.

The constraints on the chargino and neutralino sectors provide information
about the supersymmetric mass parameter $\mu$.
If the  gaugino Majorana mass parameters are unified
with a common GUT-scale mass given by $m_{1/2}$,
then the gluino, chargino and neutralino masses are
determined by $m_{1/2}$, $\mu$, and $\tan\beta$. This is the
situation in most grand unified theories (assuming that the threshold
corrections, proportional to the GUT-symmetry breaking order
parameter, are small).  In order to obtain the desired hierarchy
between the lightest chargino and neutralino masses,
our model prefers the region of parameter space where
$m_{1/2}\ll|\mu|$ (with $\mu$ negative).
Then, our choice of $M_{\widetilde\chi_1^0}\simeq 25$~GeV
fixes
\begin{eqnarray}
m_{1/2} & \simeq &  \frac{\alpha(M_{\rm GUT})}{\alpha_2(M_Z)} M_2
\simeq  110~{\rm GeV}\,.
\end{eqnarray}
In general, small threshold corrections at the grand unification
scale will modify the exact relation among the gaugino masses.
Assuming corrections of at most 25$\%$ in these relations,
the typical values for the 
masses of the other light chargino and
neutralino states are
$M_{\widetilde\chi_1^\pm}\simeq M_{\tilde\chi_2^0}\simeq 55$--70~GeV.
The dominant decay of the chargino would be 
$\widetilde\chi^+\rta\widetilde t\bar b$,
followed by $\widetilde t\rta c\widetilde\chi^0$.  Note that
the signature for this decay chain consists of bottom and charm
quark initiated jets
plus missing energy.  In particular, such events would 
be phenomenologically distinct from the usual three body decays
$\widetilde\chi^+\rta\widetilde\chi^0 W^*$ expected in the MSSM for
a light gaugino-like chargino.  The most recent data from LEP at
$\sqrt{s}=136$~GeV places severe constraints on the chargino parameters
of our model, but does not yet rule out the entire parameter space.
The choice of $m_{1/2}$ also fixes the gluino mass;
we find $M_{\tilde g}\simeq 150$--200~GeV.  
The dominant decay of this gluino would be
$\widetilde g\rta \widetilde t \; \bar t$ (or its charge-conjugated state)
followed by $\widetilde t\rta c\widetilde\chi_1^0$.
Such a gluino cannot be ruled out by present Tevatron limits.

The mass of the lightest CP-even Higgs boson should lie above the
LEP lower limit.  For $\tan\beta=1.6$, the tree-level {\it upper bound}
on the light Higgs mass is $\mhl\leq m_Z|\cos{2\beta}|\simeq 40$~GeV, 
which would
have been detected at LEP.  However, radiative corrections can
substantially raise the tree-level
upper bound \cite{radcorr}. In our model, the most
important radiative corrections are associated with the fourth
generation squark and quark masses.  We employ the radiative
corrections to the Higgs masses as outlined in
Refs. \cite{HHo,KYS,CEQR,CEQW,HHH}; it is crucial to include
the complete fourth generation squark threshold
effects, which depend on the off-diagonal squark mixing terms.
For the values of $A_{t\pri}$ and $A_{b\pri}$ obtained
from the infrared fixed point constraints and taking the $\mu$ parameter
negative and of order the soft-supersymmetry-breaking diagonal
squark mass parameters, we find that the experimental Higgs mass bound is
fulfilled even for relatively low values of the fourth generation
squark mass parameters, $m_{Q^\prime} \simgt 200$ GeV.
For example, for
$m_{Q^\prime} \simeq 250$ GeV, we find $\mhl\simeq 65$--70~GeV as a
typical range of values for the light CP-even Higgs mass, which is just
above the present LEP limits.

\section{Precision Measurements and the Top Quark Mass}

Since the fourth generation quarks are relatively heavy and
presumably are weakly mixed with the light quarks, their
contributions to electroweak observables through virtual exchange
occurs dominantly through loop corrections to the gauge boson
self-energies ({\it i.e;} the oblique corrections).  In contrast, the
third generation quark virtual effects are still responsible for
the radiative corrections to $Z \rightarrow b \bar{b}$ and
$b\rightarrow s\gamma$.  However, the
size of the virtual effects cannot be evaluated
before specifying the value of the supersymmetric Higgs
mass parameter $\mu$.

The $\mu$ parameter is constrained by the oblique corrections to
the $\rho$-parameter.  Let $\Delta\rho$ be the one-loop contribution to
$\rho$ arising from the various sectors of the supersymmetric model.
Because the top quark mass is less than half of its
standard value, the contribution of the $t$--$b$ doublet
to $\Delta\rho$ is reduced by a factor of four.
Since the fourth generation
top quark mass $m_{t\pri}$ is roughly of the same order
as the standard top one, one might expect that the fourth
generation fermions could make up the deficit in $\Delta
\rho$. However, this is not the case due to
the large value of $m_{b\pri}$.  Using
\begin{equation}
\Delta\rho(t,b)=
\frac{g^2 N_c}{64 \pi^2 m_W^2}
\left[ m_t^2 + m_b^2 - \frac{2 m_t^2 m_b^2}{m_t^2 - m_b^2}
\log\left(\frac{m_t^2}{m_b^2}\right) \right]\,,
\label{rhomumd}
\end{equation}
where $N_c=3$ is the color factor, it follows that
the fourth generation fermions contribution
to $\Delta\rho$ (for $m_{t\pri}=175$~GeV and $m_{b\pri}=105$~GeV)
is approximately the same as that from the
third generation. Hence, the supersymmetric particle contributions
must account for  roughly a half of the total
value of $\Delta\rho$.  (Explicit formulae can be found in Appendix A.)
This requirement places severe restrictions on the 
third and fourth generation squark mass
parameters $M_Q$, $M_U$ and $M_D$ [see Eq.~(\ref{stopmatrix})].
One must maximize the off-diagonal squark mixing while keeping
the diagonal squark mass parameters as small as possible.  However,
as discussed in section 4,
the latter cannot be too small; otherwise the radiative corrections to
the light Higgs mass will be reduced leading to a value of $\mhl$ below
the current LEP bound.

To obtain a sufficient squark contribution to $\Delta\rho$ is not
a simple task.  The squark contribution to $\Delta\rho$ arises only from
loops involving $\widetilde t_L$ and $\widetilde b_L$.  Since $m_b\ll 
m_Z$, there is very little mixing in the bottom squark sector, and it follows
that $M_{\widetilde b_L}\simeq M_Q$ should be larger than about
150 GeV. Furthermore, a large value of $M_Q$ implies that a significant
squark contribution to $\Delta \rho$ may only come from
large values of the off-diagonal squark mixing.
Indeed, taking into account the requirement that $M_{\widetilde t}\simeq
50$~GeV, the breakdown of the custodial $SU(2)$ symmetry is best achieved
by taking: (i) large values of the right handed squark mass parameter
($M_U \gg M_Q$) to maximize the
left handed component of the lightest stops; (ii) moderate or large values of
$M_Q$ to maximize the mass difference between the left handed
sbottom and the lightest stop; and (iii) large values
of the off-diagonal stop mixing [as required by
Eq.~(\ref{tildeat})]. The breakdown of the custodial
symmetry arises predominantly as a result of the large squark mixing
effects.  By this procedure, 
we can make the stop contribution to the $\rho$
parameter as large as desired. However, the large hierarchy between the
parameters $M_U$ and $M_Q$ and the large values of the mixing parameter
$X_t$ needed to reproduce the observed value of the $\rho$-parameter
seems rather ad-hoc.  Thus, in what follows, we focus on a more
natural choice of parameters where $M_U$ is of order $M_Q$, 
although large values of $X_t$ are required in this case as well.

There is a danger that large values
of the mixing parameter $X_t$ may render the Standard Model vacuum
state (globally) unstable \cite{stable,CW,stable2}.  In this case the true
vacuum would break electric or color charge due to a non-zero
vacuum expectation value of some squark or slepton field.  However, the
requirement of global stability of the electric and color charge
conserving vacuum is perhaps too
strong.  The Standard Model vacuum can be metastable as long as its
lifetime is sufficiently long relative to the age of the universe.
The weaker condition of vacuum metastability can permit the
large values of $X_t$ required by our model \cite{penngroup}.

The $\Delta \rho$ constraint essentially fixes the allowed range
of soft-supersymmetry breaking diagonal squark masses.  
Consider first the third generation parameters.  We take
$M_Q \simeq M_U$, where the squark mixing parameter $X_t$ is chosen in
such a way that the light stop has a mass fixed at 50 GeV.  In this case, 
the contribution of the third generation squarks to $\Delta\rho$
is too small unless
$M_Q$ lies {\it above} 200~GeV.  The larger we take $M_Q$,
the larger we must tune $A_t$ in order to preserve the light stop.
This in turn raises $\Delta\rho$.  For example, values of $m_{Q}\simeq 300$
GeV can generate a contribution from the third generation squarks which
is of the same order as $\Delta\rho(t,b)$.  However, in this case, one
must also tolerate a rather large value of $X_t \simeq 1$ TeV.

Non-negligible mixing in the fourth generation also enhances the
fourth generation squark contributions to $\Delta \rho$.  
The maximum effect is limited by the
lower bound on the mass of $\widetilde b\pri$.
In order that $t\pri\rta bW^+$ remain the dominant $t\pri$
decay, one
must kinematically forbid $t\pri\rta\widetilde b\pri\widetilde
\chi_1^+$.  Given $M_{\widetilde\chi_1^\pm}\simeq 55$--70~GeV,
a value of $M_{\tilde b\pri}\simeq 120$~GeV
is a comfortable choice.  
Since $A_{t\pri}$ and $A_{b\pri}$ are determined by
their infrared fixed point behavior, 
for a given value of the left and right handed squark mass parameters,
the value of $\mu$ which maximizes the fourth generation squark contribution
to $\Delta \rho$ can be obtained.
All the phenomenological
constraints have now forced
the parameters of the model into a very narrow corner of parameter space.

It is convenient to parameterize the oblique radiative corrections in terms
of the Peskin-Takeuchi variables \cite{peskin}
$S$, $T$ and $U$.  Here $T\equiv \alpha^{-1}
\Delta\rho$ (where $\alpha^{-1}\simeq 137$) is  the most sensitive
variable
(although some interesting restrictions can be obtained by considering $S$).
Langacker has performed a global analysis of precision electroweak
data \cite{langacker},
assuming that $m_t=80$~GeV and $\mhl=65$~GeV, and extracts values
for the oblique parameters.  He finds $T_{\rm new}= 0.70\pm 0.21$, which
in our model must arise from the contribution of the fourth
generation fermions and the third and fourth generation squarks.
(The contributions from other supersymmetric particles are negligible.)
Using Eq.~(\ref{rhomumd}),
we find that the fourth generation fermions yield a contribution of 0.25
to $T_{\rm new}$.  The contributions of the third and fourth generation
squarks depend sensitively on the squark parameters as noted above;
a range of parameters can be found that yields
a total squark contribution to $T_{\rm new}$
between 0.3 and 0.4. Due to the infrared fixed point
relation, the fourth generation mixing mass parameters
$A_{t\pri} \simeq A_{b\pri} \simeq -155$~GeV.
As an example, we find that if
%
\begin{eqnarray}
m_{Q} \simeq m_{U}
& \simeq & 275~{\rm GeV};\;\;\;\;\;\;\;\; \mu \simeq -460~{\rm GeV}
\nonumber\\
m_{Q^\prime}  \simeq m_{{U}^\prime} & \simeq &
250~{\rm GeV} \;\;\;\;\;\;\;\; A_t \simeq 695~{\rm GeV},
\end{eqnarray}
then the contribution from the third and fourth generation squarks
to the $T$ parameter is approximately equal to 0.4.
Including the rest of the supersymmetric spectrum, we find
$T_{\rm new}\simeq 0.7$, which reproduces the central
fitted value of $T_{\rm new}$ quoted above.
As for $S_{\rm new}$, its value is still subject to large
errors: $S_{\rm new} \simeq -0.16 \pm 0.21$ in the analysis of
Ref.~\cite{langacker}.  Summing the
contributions from the fourth generation
fermions and light supersymmetric particles yields a small
positive contribution to $S_{\rm new}$.  For example,
typical model parameter choices
yield a value of $S_{\rm new}\simeq 0.2$, which
lies within the $95 \%$ confidence
level bound deduced from precision electroweak data.

The exchange of third generation quarks and squarks contributes
significant radiative corrections to the $Z \rta b\bar{b}$
vertex. The fourth generation effects are suppressed by mixing
angles and are hence small. Due to the smaller top Yukawa
coupling, the radiative corrections arising from the virtual exchange of
third generation squarks are smaller than those of the MSSM for the same
choice of chargino and stop masses.
In addition, the large value of $|\mu|$ implies that
the light chargino is nearly a pure gaugino.  Thus,
even though the chargino and stop masses are close to $\half m_Z$,
the effect of the chargino-stop loop has only
a small effect on the rate  $Z \rta b\bar b$.
Our model then predicts 
\begin{equation}\label{modelrb}
R_b \simeq 0.2184, 
\end{equation}
which, under the assumptions specified in
section 2, is within one standard deviation of the measured
LEP value [Eq.~(\ref{zbbnumber2})].
The improvement over the Standard Model prediction
is mainly due to the fact that $m_t\simeq m_W$.

The value of $R_b$ has also consequences for the value of $\alpha_s(M_Z)$
extracted at LEP.  The value of $R_\ell$,
from which $\alpha_s(M_Z)$ at LEP is extracted,
would be different from the Standard
Model value, due to the modification of $\Gamma_{\rm had}$ induced through
the larger value of $\Gamma_b$. In models where only the bottom-quark
width of the $Z$-boson is modified by virtual corrections,
a simple relation between $\alpha_s(M_Z)$ and $R_b$ may be obtained.
In the one-loop approximation, it is given by
\begin{equation}
{\alpha_s(M_Z)\over\pi} = \left({1-R_b\over 1-R_b^0}\right)
\left({\alpha_s^{0}(M_Z)\over\pi}\right)-
 \frac{R_b - R_b^0}{1-R_b^0},
\label{alphas}
\end{equation}
where $\alpha_s^0(M_Z) = 0.124\pm 0.005$ is the value that would be obtained
from the measured value of $R_\ell$ in the Standard Model
(assuming $\mhl\simeq 60$~GeV), and $R_b^0 = 0.2156$
is the Standard Model value for $R_b$.
Using the value of $R_b$ predicted by our model [Eq.~(\ref{modelrb})],
we find $\alpha_s(M_Z) \simeq 0.112 \pm 0.005$, in good agreement with
determinations of $\alpha_s$ from deep inelastic scattering data
\cite{hinchliffe}.

\section{Flavor Changing Neutral Current Processes}

The low value of the top quark mass can have far reaching consequences
for flavor changing neutral current (FCNC) processes. We shall analyze
the predictions for three particularly sensitive FCNC observables:
the $b \rightarrow s \gamma$ decay rate, the
$B_d$--$\overline{B}_d$ mixing and the CP-violation 
parameter $\epsilon_K$ \cite{bsga}. 
In our four generation supersymmetric model,
the FCNC amplitudes depend not only on the top quark mass but
also on the full set of supersymmetric parameters. In particular,
the contribution of light stops and light charginos to these amplitudes 
can be quite significant. Moreover, the predictions depend strongly
on the assumptions about the flavor structure of the soft-supersymmetry-%
breaking parameters. As an indicative example, we shall assume that a
super-GIM mechanism is active.  In addition, we further assume that the
mixing of the fourth generation with the first and second generations
is negligibly small.
We shall present the results for $M_{\widetilde{\chi}^\pm} \simeq
65$~GeV (the amplitudes considered below are slightly
reduced for larger values of the chargino mass).  The other model
parameters are fixed by the phenomenological considerations of
the previous sections.

One of the
most sensitive tests of the model is to check that its prediction for
$b\rta s\gamma$ is consistent with $1.0\times 10^{-4}\simlt
{\rm BR}(b\rta s\gamma)\simlt 4\times 10^{-4}$, as required by the CLEO
measurement \cite{cleo}. In Figure 4 we plot the prediction for
the branching ratio ${\rm BR}(b\rta s\gamma)$ in our model.
It follows  that, for values of the CP-odd Higgs mass 
$m_A \simgt 100$~GeV, the prediction of our model lives comfortably
within the experimental bound.

\begin{figure}[htbp]
\centering
\centerline{
\psfig{file=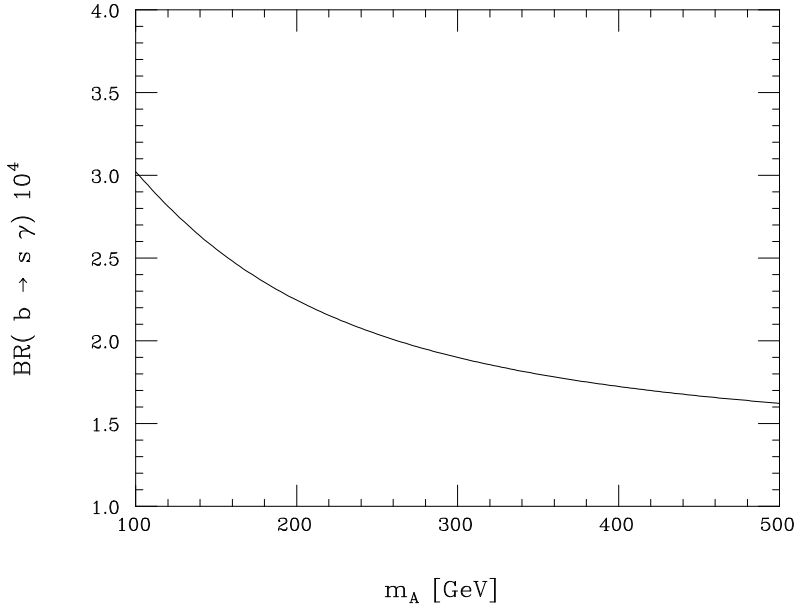,height=8cm,angle=90}}
\caption[0]{
Branching ratio ${\rm BR}(b\rightarrow s\gamma)$ as a function of
the CP-odd mass $m_A$.}
\end{figure}

The bounds on $B_d$--$\overline{B}_d$ mixing and $\epsilon_K$ 
can place a lower
bound on the Standard Model top quark mass value.
This bound can be explicitly obtained from the following 
two equations taken from Ref.~\cite{Buras}.
From the measurement of $\epsilon_K$, one deduces that:
\begin{equation}
\eta \left[ ( 1- \rho) A^2 \eta_2 \; P + P_c \right] A^2 B_K \simeq 0.223\,,
\end{equation}
where $\eta$ and $\rho$ are parameters arising in the Wolfenstein
parameterization of the CKM matrix (which are constrained by the
relation $\sqrt{\rho^2+\eta^2}=0.355\pm 0.090$), 
$A\equiv |V_{cb}|/|V_{us}|^2$ is found from data to be $A=
0.826\pm 0.060$, $\eta_2\simeq 0.57$ is a QCD
correction factor, $P$ is a reduced amplitude that arises in the
computation of the $W$--$t$ box diagram, $P_c\simeq 0.3$ is a small
correction that arises from the box diagram mediated by the charmed
quark, and $B_K=0.8\pm 0.2$ represents the uncertainty in the hadronic
matrix element.   

From the measurement of $B_d$--$\overline{B}_d$ mixing, one deduces that: 
\begin{equation}
\left|{V_{td}\over V_{us}V_{{cb}}}\right|^2
\left({\frac{0.75}{x_d}}\right) \left({\frac{F_B \sqrt{B_B}}
{200~{\rm MeV}}}\right)^2 \left|{\frac{V_{cb}}{0.040}}\right|^2 
\left({\frac{\tau_B}{1.6~{\rm ps}}}\right)^2 P\simeq 2.089\,,
\end{equation}
where $F_B\sqrt{B_B}= 200\pm 40$~MeV, $|V_{cb}|= 0.040\pm 0.003$, 
$\tau_B=1.6\pm 0.1$~ps, and $x_d\equiv \Delta M_B/\Gamma_B= 0.75\pm 0.06$.

In the Standard (non-supersymmetric three-generation) Model, the
reduced amplitude $P$ depends on the top quark mass according to the
following approximate formula:
\begin{equation}
P_{\rm SM}\simeq 0.79 \left( \frac{{M}_t}{m_W} \right)^{1.52},
\end{equation}
where $M_t\equiv m_t(m_t)$ is the $\overline{\rm MS}$ running top quark mass
evaluated at the pole mass.  For example, for a running
mass $M_t = 170$ GeV, $P\simeq 2.465$.
Scanning over the allowed values of all free parameters at the
one sigma level we find
\begin{equation}
0.8 \simlt \; P \; \simlt 10.5\,.
\label{ekbounds}
\end{equation}
The lower bound on $P$ is relevant for our discussion.
In particular, the lower bound on $P$ translates into a lower
bound on the top quark mass given by $M_t\simgt 80$~GeV.



\begin{figure}[ht]
\centering
\centerline{
\psfig{file=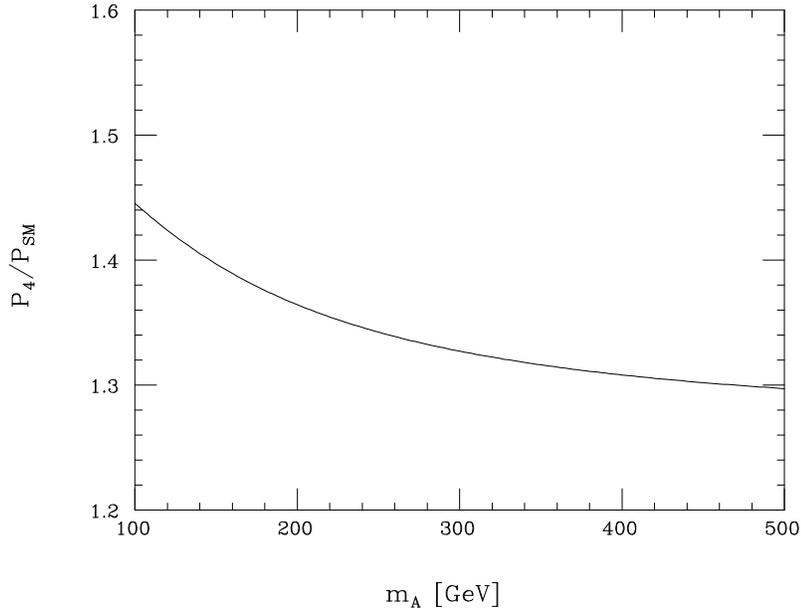,height=8cm,angle=90}}
\caption[0]{ Ratio
of the $B_d$-$\overline{B}_d$ reduced 
amplitude $P_4$ in the four generation model to the corresponding
amplitude $P_{\rm SM}$ in the Standard Model,  
for $m_t=80$~GeV as a function of the
CP-odd Higgs mass $m_A$.}
\label{ratiobma}
\end{figure}

We can now apply this analysis to any model for which an effective 
GIM mechanism is active.  In particular, let $P_4$ be the value of $P$
obtained in our four-generation low-energy supersymmetric model.
We find that $P_4$ is
enhanced with respect to $P_{\rm SM}$ for the same
top quark mass, primarily due to the effects of light supersymmetric
particle exchange.
Figure~\ref{ratiobma} shows the ratio $P_4/P_{\rm SM}$ as a function of
the CP-odd Higgs mass.
We see that the enhancement factor is approximately
1.35, with a dependence on $m_A$ an indication of the size of the 
charged Higgs exchange.  Hence, the reduced amplitude in our model is 
$P_4 \simeq 1.01$;
which is within the one sigma bound quoted above.  Note that
this value strongly depends on our assumption about
the presence of an effective GIM mechanism. 
If the model contains extra
flavor violation parameters, the prediction for $P_4$ can
be significantly modified. Furthermore, if new
CP-violation phases are present in the theory, the above bounds
on $P_4$ [Eq.~(\ref{ekbounds})] can be further relaxed.

\newpage

\section{Discussion and Conclusions}
\medskip

A low energy supersymmetric model with four generations
has been presented with $m_t \simeq m_W$, which satisfies
all present day phenomenological constraints.
Typical values of the model spectrum are shown in Table 1.
The fourth generation quark $t\pri$ is identified
as the state discovered by the CDF and D0 Collaborations, while the
light top quark decays via $t\to\widetilde t\widetilde
\chi_1^0$, thereby escaping previous searches at hadron colliders.
The oblique radiative corrections inferred from precision electroweak
measurements at LEP are reproduced by contributions
from the third and fourth generation quarks and squarks.
The most
theoretically troubling feature of the model is the large mixing among
the third generation squarks that is necessary to ensure a light
top squark and provide a viable
prediction for the electroweak $\rho$-parameter.

\begin{table}[htbp]
\begin{center}
\caption{Example of Model Spectrum and Parameters
               ($\tan\beta=1.6$)}
\vskip1pc
\renewcommand{\arraystretch}{1.2}
\begin{tabular}{|lc|lc|}
\hline
Particle&  Mass (GeV) & Particle & Mass (GeV) \\
\hline
$t$ &     \980 &    $\widetilde\chi^0_1$ & \925 \\
$b\pri$& 105 &   $\widetilde\chi^+_1$ &  \965 \\
$t\pri$& 175 &   $\widetilde\chi^0_2$ & \966 \\
$\tau\pri$& \950 & $\tilde g$ & 190 \\
$N_{\tau\pri}$& \980& $\widetilde b$ & 264 \\
$h^0$&      \967&      $\widetilde t$ & \950 \\
$A^0$&       300&      $\widetilde b\pri$ & 116 \\
$H^\pm$ &  310&      $\widetilde t\pri$ & 263 \\
\hline\hline
Parameter & Value (GeV) &   Parameter & Value (GeV) \\
\hline
$\mu$&  $-460$ &   $M_{\widetilde Q,\widetilde U,\widetilde D}$ & \9275 \\
$M_1$& \9\9 23 &   $M_{\widetilde Q\pri,\widetilde U\pri,\widetilde
D\pri}$ & \9250 \\
$M_2$& \9\9 54 &   $A_{t,b}$ & \9695 \\
$M_3$&  \9 170 &   $A_{t\pri,b\pri}$ & $-155$ \\[2pt]
\hline
\end{tabular}
\end{center}
\end{table}

There are several remarkable properties of the present
model in contrast to the MSSM which should be stressed: 

1. In the MSSM, any physics leading to larger values of $R_b$
also contributes to non-standard top quark decays \cite{topdec}.
Indeed, the potentially viable
scenarios invoke either light charginos\footnote{Based on the absence of
light charginos in the most recent LEP run at $\sqrt{s}=136$~GeV,
Ref.~\cite{jellis} quotes an absolute upper limit of $R_b<0.2174$ for the
three-generation MSSM in the case of small $\tan\beta$.}
and light stops for small values of $\tan\beta$
and/or light charged Higgs bosons
for large values of $\tan\beta$. Light charginos are usually associated
with light neutralinos; hence, the decay mode
$t \rightarrow \widetilde{t} \widetilde{\chi}^0$ is present, which can
account for a significant fraction of the top quark branching ratio.
As a result, in such models, the
typical branching ratio of non-Standard Model top quark decays
is of order 30\%.   
This is in slight disagreement with the Tevatron
top decays observations, although the experimental errors are still large
enough to permit a non-negligible reduction of BR$(t\to bW^+)$ from 1.
In our model, the decay of $t\pri \rightarrow \widetilde{t}
\widetilde{\chi}^0$ is a flavor violation process, which only
occurs at the one loop level, and hence is naturally suppressed 
relative to the decay $t\pri \rightarrow b W^\pm$.

2. The large value of the top quark mass in the MSSM
tends to lead to FCNC processes (which are particularly striking
in the leptonic sector) induced by
physics at the grand unification scale \cite{gutfcnc}.
In our case,
these effects are reduced due to the lower value of
the top quark Yukawa coupling at $M_{\rm GUT}$ and the
expected small mixing between the fourth and the first two
generations.

3. The model presented in this paper exhibits the
unification of the four fourth generation fermion-Higgs
Yukawa couplings at a common high energy scale.
In the MSSM, the unification of the bottom and top quark Yukawa
couplings can only be obtained for very large values of
$\tan\beta$, which usually implies a large fine tuning of
the Planck scale parameters \cite{largetanb}.
In contrast, $\tan\beta\sim{\cal O}(1)$ in our model, which is
a more natural value for a model of bottom--top Yukawa coupling
unification (and is certainly less fine-tuned than the large $\tan\beta$
model), especially given the proximity of the Yukawa couplings
to their infrared fixed point values.


In a complementary work, Gunion, McKay and Pois \cite{pois}
constructed four-generation models in the framework of minimal
low-energy supergravity.
The state discovered by the CDF and D0 Collaborations near 175~GeV
in mass was taken to be the top quark.  Thus, to keep the fourth
generation quark-Higgs
Yukawa couplings perturbative up to the GUT scale, one
must hide the $b\pri$ and $t\pri$ in a mass region
below $m_t\simeq 175$~GeV.   In this alternative approach, the $R_b$
anomaly would have to be explained by assuming an appropriately light
super-partner spectrum (as in the case of the three-generation MSSM).
Gunion {\it et al.} also performed a complete renormalization group
analysis of the soft-supersymmetry breaking terms in the four-generation
models.  In particular, they demonstrated that our model
cannot be obtained from universal boundary conditions 
for the soft-supersymmetry-breaking parameters at $M_{\rm GUT}$.
This result was due primarily to the large values of
$A_t(M_{\rm GUT})$ required to achieve a light top squark.  Although
helpful in suppressing flavor changing processes, the universality of
scalar masses is not a necessary condition
for obtaining an acceptable phenomenology.
Even in the case of the three-generation MSSM, 
the violation of the universal boundary conditions
seems to be a requirement for obtaining values of $R_b$ closer to the
present experimental values \cite{Gordy,CW,talk}.

We have demonstrated that present day experimental data place severe
constraints on the four-generation model parameters.  Moreover, the
model possesses a rich spectrum of new particles that will be accessible
to LEP-2 and the Tevatron.   In particular, eight new particles of this
model could be discovered at LEP-2: the $t$-quark, the fourth generation
leptons ($\tau\pri$ and $N_{\tau\pri}$), the light Higgs
boson ($\hl$), and four supersymmetric particles ($\widetilde\chi_1^0$,
$\widetilde\chi_2^0$, $\widetilde\chi_1^\pm$, and $\widetilde t$).
New bounds coming from 
the initial run of LEP-2 at $\sqrt{s}=136$~GeV have constrained the
model further.  For instance, non-observation
of the $\tau\pri$ requires that it must be
almost degenerate in mass with the fourth generation neutrino.
Moreover, the non-observation of the lightest chargino pushes its
mass to its upper allowed range, $M_{\widetilde{\chi}^\pm} \simeq$~65--70
GeV.  One can imagine scenarios where some of the properties of the
eight particles listed above are altered to avoid conflict with future
collider limits.\footnote{As an example, if gaugino masses are not unified
then the mass of $\widetilde\chi^\pm_1$ is independent of the mass
of the lightest neutralino if both states are dominantly gaugino-like
({\it e.g.}, assuming $|\mu|$ is large compared to the gaugino mass
parameters).  Then, by choosing the $\widetilde\chi_1^\pm$ mass above
100~GeV, one can tolerate a substantially lighter $\widetilde b^\prime$
than shown in Table 1, while still kinematically forbidding the
two-body decay $t^\prime\to \widetilde b^\prime\widetilde\chi_1^+$.
(A lighter $\widetilde b^\prime$ can be obtained
by adjusting $\mu$ to generate larger mixing
among the fourth generation squarks.)
Three ``benefits'' ensue: (i) a lighter $\widetilde b^\prime$ can generate
a larger fourth generation squark contribution to $\Delta\rho$
and increase the radiatively corrected value of $m_{h^0}$, thus allowing
for less extreme mixing parameters in the third generation squark sector;
(ii) the $b^\prime$ quark would be more difficult to detect at the
Tevatron, since in this case $b^\prime\to\widetilde b^\prime
\widetilde\chi_1^0$ followed by $\widetilde b^\prime\to b\widetilde
\chi_1^0$ would be the dominant decay chain; and (iii) a heavier
chargino would be kinematically inaccessible to LEP-2.}
However, it is certain that our model is ruled out
unless a light top squark is discovered in LEP-2 data expected to be
collected in 1996, or perhaps in the Run Ib data from the Tevatron
after the complete data samples have been analyzed.  Of course, if 
such a state is discovered, it could be consistent with the
three-generation MSSM.  In this case, the light top quark in our
approach (or a fourth generation quark doublet with mass below that of
the top quark in the approach of Gunion {\it et al.}) must be found to
confirm the existence of a new generation.  Alternatively,  
the exclusion of these models would raise ones confidence
that the number of families in the effective low-energy supersymmetric
theory below $M_{\rm GUT}$ is indeed three.

\vspace*{2pc}
\centerline{\bf Acknowledgements}
\vspace*{1pc}
We would like to thank Paul Langacker and Jens Erler for running their
precision electroweak programs with $m_t\simeq m_W$, and for 
enlightening conversations.  We are grateful to
Ahmed Ali for encouraging us to take a more careful look at
the implications of $\epsilon_K$ and $B$--$\overline B$ mixing, and
to Guido Altarelli, George Hou, Mathias Neubert, and 
Sandip Pakvasa for a number of useful remarks.
One of us (HEH) would like to acknowledge the support of the
U.S. Department of Energy.
\vskip 1cm
\appendixA{Appendix A~~~Squark contributions to the $\rho$-parameter}

This appendix contains the formulae for the squark contributions
to the $\rho$-parameter.  Only the contribution of the third
generation squarks will be written out explicitly below.  The
contributions of the other squark generations can be trivially
obtained by employing the appropriate squark masses and mixing
angles (to a good approximation, squark mixing is appreciable only in
the third and fourth generations).
The formulae below can also be applied to the slepton
contributions by setting the color factor ($N_c=3$ for squarks) equal
to unity.
 
We define the squark mixing angles as follows.  Let $\widetilde q_1$
and $\widetilde q_2$ be the quark mass eigenstates,
with $M_{\tilde q_1}< M_{\tilde q_2}$.  (In this 
paper, we have denoted the light squark eigenstate 
$\widetilde q_1$ by $\widetilde q$, since the heavier squark 
eigenstate does not play an important role in our considerations.)
Then, in terms of the
interaction eigenstates $\widetilde q_L$ and $\widetilde q_R$,
\begin{eqnarray}\label{squarkangles}
\widetilde q_1 & = & -\widetilde q_L\sin\theta_q+\widetilde
q_R\cos\theta_q\,,\nonumber \\ 
\widetilde q_2 & = & \widetilde q_L\cos\theta_q+\widetilde
q_R\sin\theta_q\,. 
\end{eqnarray}
For example, after diagonalizing the top-squark
mass-squared matrix [Eq.~(\ref{stopmatrix})], one obtains
\begin{equation}\label{tantheta}
\tan 2\theta_t= {2m_t(A_t-\mu\cot\beta)\over
M_Q^2-M_U^2+(c_L-c_R)m_Z^2}\,.
\end{equation} 

The contribution of the third generation squarks to the
$\rho$-parameter is then given by:
\begin{eqnarray}\label{rhoparm}
\Delta\rho(\widetilde t,\widetilde b) & = & {g^2 N_c\over 32\pi^2m_W^2}
\Biggl[\sin^2\theta_t\sin^2\theta_b F(m^2_{\tilde t_1},m^2_{\tilde b_1})+
\sin^2\theta_t\cos^2\theta_b F(m^2_{\tilde t_1},m^2_{\tilde b_2})
\nonumber \\
&&\qquad\qquad +
\cos^2\theta_t\sin^2\theta_b F(m^2_{\tilde t_2},m^2_{\tilde b_1})+
\cos^2\theta_t\cos^2\theta_b F(m^2_{\tilde t_2},m^2_{\tilde b_2})
\nonumber \\
&&\qquad\qquad
-\sin^2\theta_t\cos^2\theta_t F(m^2_{\tilde t_1},m^2_{\tilde t_2})
-\sin^2\theta_b\cos^2\theta_b F(m^2_{\tilde b_1},m^2_{\tilde
b_2})\Biggr]
\nonumber \\
\end{eqnarray}
where the function $F$ is defined by
\begin{equation}\label{functionf}
F(m_1^2,m_2^2)\equiv\half(m_1^2+m_2^2)-{m_1^2 m_2^2\over m_1^2-
m_2^2}\,\ln\left({m_1^2\over m_2^2}\right)\,.
\end{equation}

Note that in the supersymmetric limit ($M_{\tilde t_1}=M_{\tilde t_2}=m_t$ and
$M_{\tilde b_1}=M_{\tilde b_2}=m_b$), 
one finds $\Delta\rho(\widetilde t,\widetilde b)=
\Delta\rho(t,b)$ [see Eq.~(\ref{rhomumd})].
In contrast, in the limit where $M_Q^2\gg m_Z^2$,
$m_t^2$, the squark contribution to the $\rho$-parameter decouples:
\begin{equation}\label{rhodecouple}
\Delta\rho(\widetilde t,\widetilde b)\simeq  {g^2N_c m^4_t K\over
   32\pi^2m^2_W M_Q^2}\,,
\end{equation}
where $K$ approaches a constant in the limit of large $M_Q^2$.
One can evaluate $K$ by using the following properties of
$F(m_1^2,m_2^2)$ in the
limit of $|m_1^2-m_2^2|\ll m_1^2$, $m_2^2$
\begin{eqnarray}\label{fprops}
F(m_1^2,m_2^2) & \simeq & {(m_1^2-m_2^2)^2\over 6m_2^2}\,, \\
F(m_1^2,m_3^2)-F(m_2^2,m_3^2) & \simeq & (m_1^2-m_2^2)
  \left[{1\over 2}+ {m_3^2 \over m^2_3-m^2_2}
  + {m^4_3\over (m^2_3-m^2_2)^2} \ln\left({m^2_2\over m^2_3}\right)
  \right]\,.\nonumber
\end{eqnarray}
In this paper, it is the important third and fourth generation
squark mixing effects that lead to a significant supersymmetric
contribution to the $\rho$-parameter. 



\end{document}